\documentclass[aps,prd,twocolumn,groupedaddress]{revtex4}

\usepackage{graphicx}

\begin{document}


\title{$^6$Li Production by the Radiative Decay of Long-Lived Particles}


\author{Motohiko Kusakabe}
\email[]{kusakabe@th.nao.ac.jp}
\thanks{Research Fellow of the Japan Society for the Promotion of Science}
\author{Toshitaka Kajino}
\affiliation{Department of Astronomy, School of Science, University of
Tokyo,  Hongo, Bunkyo-ku, Tokyo 113-0033, Japan\\
Division of Theoretical Astronomy, National Astronomical
Observatory of Japan, Mitaka, Tokyo
  181-8588, Japan}

\author{Grant J. Mathews}
\affiliation{Center for Astrophysics, University of Notre Dame, Notre
Dame, IN 46556, U.S.A.}


\date{\today}

\begin{abstract}

Recent spectroscopic observations of metal poor stars have indicated that both
$^7$Li and $^6$Li have abundance plateaus with respect to the
 metallicity.  Abundances of $^7$Li are about a factor
three lower than the primordial abundance predicted by standard big-bang
 nucleosynthesis (SBBN), and $^6$Li
abundances are $\sim 1/20$ of $^7$Li, whereas SBBN predicts negligible
amounts of $^6$Li compared to the detected level.  These
discrepancies suggest that $^6$Li has another cosmological or Galactic
origin than the SBBN.  Furthermore, it would appear that $^7$Li (and also $^6$Li) has been depleted from its primordial abundance by some post-BBN processes.  In this paper we study the possibility that the radiative decay of long-lived particles
has affected the cosmological lithium abundances.  We calculate the
non-thermal nucleosynthesis associated with the radiative
decay taking account both of the primary nuclear production reactions and the effects of secondary production as well as the 
destruction processes of
energetic nuclides D, T, $^3$He, $^4$He, $^6$Li and $^7$Li.  We explore the allowed region of the parameters
specifying the properties of long-lived particles.  We also impose 
constraints from observations of the CMB energy spectrum. 
We find that non-thermal
nucleosynthesis produces $^6$Li at the level detected in metal poor halo
stars (MPHSs), when the lifetime of the unstable particles is of the order $\sim
10^8-10^{12}$~s and their initial abundance with respect to that of the
photons is $\sim (10^{-13}-10^{-12}~{\rm GeV})/E_{\gamma 0}$, where
$E_{\gamma 0}$ is the emitted photon energy in the radiative decay.  We
conclude that the most likely nucleosynthetic scenario involves two different processes.  
First, a non-thermal
cosmological nucleosynthesis associated with the radiative decay
of unstable particles; and second, the stellar depletion of both of the
primordial lithium isotopic abundances.
\end{abstract}



\maketitle

\section{Introduction}
In standard cosmology, the universe is thought to have experienced big-bang nucleosynthesis
(BBN) at a very early stage.  D, T, $^3$He, $^4$He,
$^6$Li, $^7$Li and $^7$Be are produced appreciably at this
epoch.  Different types of observations have been made for the purpose of
determining the primordial elemental abundances.  These observational results
provide rich information about the cosmic chemical evolution of the
light nuclear
species.  Today BBN provides a very precise tool for
inferring the condition of the early universe.  BBN in the standard cosmology explains relatively well the inferred
primordial abundances for a narrow range of the universal
baryon-to-photon ratio $\eta$.

The Wilkinson Microwave Anisotropy Probe (WMAP) satellite has measured
the temperature fluctuations of the cosmic microwave background (CMB)
radiation, and parameters characterizing the standard big bang 
cosmology have been deduced~\cite{Spergel:2003cb,Spergel:2006hy} from
these data.  For the baryon-to-photon ratio $\eta$ deduced from fits to the CMB, 
the BBN model predicts abundances of the light elements which are more-or-less consistent
with those inferred from astronomical observations.  This near agreement places
significant limits on 
non-standard models which influence the cosmic nuclear abundances.  

In this regard, unstable massive particles decaying during or after the BBN epoch are
strongly constrained~\cite{ellis85,Reno:1987qw,dimopoulos88,khlopov94}.  For example, radiative decay~\cite{Jedamzik:1999di,Kawasaki:2000qr,Cyburt:2002uv},
 hadronic decay, or
annihilation~\cite{Jedamzik:2004er,Jedamzik:2004ip,Kawasaki:2004qu} have
been studied recently, and several critical constraints on the relic
particle properties were derived.  These particle processes
induce electromagnetic and/or hadronic showers which lead to the destruction
of preexisting nuclei and to the production of
different nuclear species.  In turn, these modifications to the light
element abundances are used to constrain theories for the
decay of relic particles.  The physics associated with the general decay
process (including a hadronic decay component)
 is explained in~\cite{Kawasaki:2004qu,Jedamzik:2006xz}.

Super-weakly interacting
massive particles (SWIMPs) have been proposed~\cite{Feng:2003xh} as
candidates for cosmological dark matter.  They are a form of
non-baryonic dark matter derived from theories beyond the standard model
(SM) such as
supersymmetry or universal extra dimensions (UED)~\cite{Servant:2002aq}.  
The stable particles in these
theories may constitute the dark matter, while the unstable particles could decay
to modify the nuclear abundances in the early universe.  This type of dark-matter
particle interacts
super-weakly with the SM fields and cannot be detected directly in
experiments searching for conventional
WIMPs.  Thus, further studies of decay-induced nucleosynthesis could help
to constrain such new particle theories.

Spectroscopic lithium abundances have been detected in the atmospheres
of metal poor stars.  Nearly constant
abundances of $^6$Li and $^7$Li in metal-poor
Population II (Pop~II) stars have been
inferred~\cite{Ryan:1999vr,Melendez:2004ni,Rollinde:2004kz}.  Spectroscopic measurements 
obtained with high resolution 
indicate that metal poor halo stars (MPHSs) have a very large abundance of $^6$Li, i.e.~at
a level
of about a twentieth that of $^7$Li.  This is about three orders of magnitude
larger than the SBBN prediction of the $^6$Li abundance.  

It was
suggested some time ago~\cite{dimopoulos88} that a high $^6$Li abundance could be produced via particle decay.  More recently,
radiative particle decay in particular has been proposed~\cite{Jedamzik:1999di} as the source of the abundant
$^6$Li observed in low metallicity stars.  Furthermore, it was suggested~\cite{Jedamzik:2004er} that
hadronic decays which occur at around 1000 seconds after the big-bang could simultaneously
resolve both the $^6$Li and $^7$Li abundance issues.  This probability has been studied within the context of
supersymmetry, and there is sufficient parameter space for this to be possible~\cite{Jedamzik:2005dh}.  A
previous study on the radiative decay, however, was unable to provide a
thorough solution to the lithium abundance puzzle~\cite{Ellis:2005ii}.  In this paper, therefore,
we consider this issue to be as of yet unresolved.  We independently 
calculate the nucleosynthesis triggered by the radiative decay
processes of long-lived relic particles over a wide range of parameters specifying the properties of 
the relic particles.  We take into account 
the primary, secondary, and tertiary processes resulting from the
electromagnetic cascade showers which both produce and destroy the light
elements.  We then constrain the abundance of long-lived particles
from the calculated nucleosynthesis.  We do not find, however, a simultaneous solution to both
the $^7$Li and $^6$Li abundances unless there is stellar destruction of lithium.

This paper consists of the following structure.  In Sec.~\ref{sec2} we
describe briefly the formulae of non-thermal
nucleosynthesis including the relevant photo-dissociation and nucleus-nucleus
collisions.  In Sec.~\ref{sec3} we summarize the observed light element
abundances and the constraints adopted in this work.  The calculated
results are shown in Sec.~\ref{sec4}, where we derive the constraints on the
parameters of decaying particles.  In Sec.~\ref{sec5}, the constraints
from the CMB spectrum are
imposed, and the effects on small scale structure are analyzed.  The most
interesting parameter region is deduced by taking the observational constraints on the light element
abundances into account.  We conclude that our model can explain
the desired $^6$Li production by non-thermal nucleosynthesis even if there is stellar destruction of
both lithium isotopes to explain the observed $^7$Li.

\section{Model}\label{sec2}
We assume the creation of high energy photons from the radiative decay of
a massive particle with a lifetime of 10$^{2}$~-~10$^{12}$~s.  In this
section we describe equations to calculate the non-thermal nucleosynthesis
triggered by the appearances of high energy
photons from decay.  See~\cite{Cyburt:2002uv} for details on the formulation which we
adopt.
\subsection{Initial non-thermal photon spectrum}
We assume that the decaying dark particle is
non-relativistic, and almost at rest in the expanding
universe.  We denote the imaginary particle by $X$, [e.g. corresponding to
the next-to-lightest
supersymmetric particle (NLSP)] with a mass $M_X$ that
decays into
a photon plus another dark-matter particle, [e.g. the lightest supersymmetric
particle (LSP)].  We represent the emitted photon energy by $E_{\gamma
0}$.

When an energetic photon emerges, it interacts with
the cosmic background and induces an electromagnetic cascade shower.  The
faster processes are pair production through background photons
$\gamma_{\rm bg}$ ($\gamma
\gamma _{\rm bg}\rightarrow e^+ e^-$) and inverse Compton scattering of
produced electrons and positrons through background photons ($e^\pm
\gamma _{\rm bg}\rightarrow e^\pm \gamma$).  These two processes produce electromagnetic showers and the
non-thermal photon spectrum realizes
a quasi-static equilibrium~\cite{Kawasaki:1994sc,Protheroe:1994dt}.  The attained zeroth
generation photon spectrum can be written~\cite{berezinskii90},
\begin{equation}
{p_{\gamma}(E_{\gamma})}
\approx \left\{ \begin{array}{ll}
K_0(E_X/E_{\gamma})^{1.5}
&  \mbox{for $E_{\gamma}<E_X$} \\ 
K_0(E_X/E_{\gamma})^{2.0}
&  \mbox{for $E_X\leq E_{\gamma}<E_C$}, \\ 
0
&  \mbox{for $E_C\leq E_{\gamma}$} \\ 
\end{array}
\right.	
\label{spectrum}
\end{equation}
where $K_0=E_{\gamma 0}/\left(E_X^2\left[2+\ln(E_C/E_X)\right]\right)$ is a
normalization constant fixed by energy conservation of the
injected photon energy.  This spectrum has a break in the
power law at $E_\gamma=E_X$ and an upper cutoff at $E_\gamma=E_C$.  We take
the same energy scaling with the temperature $T$ of the background
photons as in~\cite{Kawasaki:1994sc}, i.e.~$E_X=m_e^2/80T$ and $E_C=m_e^2/22T$, where the cascade
spectrum was calculated by numerically solving a set of Boltzmann
equations.  

Above $E_C=m_e^2/22T$, the rapid
interaction between photons and electrons or positrons causes almost all
energetic photons to quickly lose their energy.  This leaves
effectively no number spectrum for energies $E_\gamma> E_C$.  In the context
of the theory of electromagnetic cascade showers~\cite{berezinskii90},
the scale $E_X=m_e^2/80T$ corresponds to the point where the
energy of the energetic photon is
below the threshold for double
photon pair creation.  Hence, pair creation from interactions between 
energetic photons and the cosmic background radiation (CBR) ceases to operate.

The zeroth-generation non-thermal photons experience additional processes including:
Compton scattering ($\gamma e^\pm_{\rm bg} \rightarrow
\gamma e^\pm$); Bethe-Heitler ordinary pair creation
in nuclei ($\gamma N_{\rm bg}\rightarrow e^+ e^- N$); and double photon
scattering ($\gamma \gamma_{\rm bg} \rightarrow \gamma \gamma$).  These
slower processes further degrade the quasi-static equilibrium
photon spectrum.

Because the rates of these electromagnetic interactions are faster than
the cosmic expansion rate, the photon spectrum is modified into a new quasi-static equilibrium
(QSE).  This distribution is given by
\begin{equation}
{\mathcal N}_\gamma^{\rm QSE}(E_\gamma) =
\frac{n_Xp_\gamma (E_\gamma)}{\Gamma_\gamma(E_\gamma)\tau_X},
\label{ng}
\end{equation}
where $n_X=n_X^0(1+z)^3 \exp(-t/\tau_X)$ is the number
density of the decaying particles at a redshift z, and $\tau_X$ is its mean life.
The quantity $\Gamma_\gamma$ is the energy degradation rate through the three slower processes for the
zeroth-generation photons.  We use this steady state approximation for the cosmic
non-thermal constituent of photons.  The
Boltzmann equation describing the electromagnetic cascade processes
is compiled in~\cite{Kawasaki:1994sc}.  We used their relevant reaction rates
 to compute the cascade shower.

Because the cutoff
scale of the photon energy is inversely proportional to the background
temperature $T$, the cutoff scale comes to large as the universe cools.
As a result, the number of the energetic non-thermal photons increases at
low $T$.

\subsection{Primary nucleosynthesis}
The equation for the production and destruction of nuclei by
non-thermal photons is given by 
\begin{eqnarray}
\frac{d Y_A}{d t} &
\hspace{-5pt}
=&
\hspace{-5pt}
 \sum_T \frac{N_A(T)}{N_T(T) !} Y_T \int_{0}^\infty
\hspace{-5pt}
 dE_\gamma
{\mathcal N}_\gamma^{\rm QSE}(E_\gamma)\, 
\sigma_{\gamma + T\rightarrow A} (E_\gamma)
\nonumber \\
&&
\hspace{-5pt}
- Y_A\sum_P
 \frac{N_A(P)}{N_A(P) !}\int_{0}^\infty 
\hspace{-5pt}
dE_\gamma {\mathcal N}_\gamma^{\rm QSE}(E_\gamma)\, 
\sigma_{\gamma + A\rightarrow P} (E_\gamma),
\nonumber \\
\label{dydt}
\end{eqnarray}
where $Y_i\equiv n_i/n_B$ is the mole fraction of a particular nuclear
species $i$, and $n_i$ and $n_B$ are number densities of
nuclei $i$ and baryons.  The first and second term on the
right-hand side are the source and sink terms for nucleus $A$.  The source
term includes the creation of nuclide $A$ from all possible target nuclides
$T$ through a
non-thermal photo-dissociation reaction $\gamma + T \rightarrow A$.  And the sink term
includes the destruction of nuclide $A$ that is specified by the reaction
$\gamma + A \rightarrow P$ for any produced nuclides $P$.  $E_\gamma$ is
a non-thermal photon energy.  The cross sections of processes $\gamma
+ T \rightarrow A$ and $\gamma + A \rightarrow P$ are denoted by $\sigma_{\gamma
+ T \rightarrow A}(E_\gamma)$ and $\sigma_{\gamma+ A \rightarrow
P}(E_\gamma)$.  Further we use $N_K(L)$ to represent the number of a particular particle $K$
relevant to a process $\gamma+a\rightarrow b$, where $K$ is either $a$ or
$b$ and $L$ is $T$ or $P$.  For example, in the process
$^4$He($\gamma$,$d$)$d$, $N_d$($^4$He)=2 because two deuterons are
produced from $^4$He.

Next, we formulate the rate equation.  We define $r\equiv n_X^0/n_\gamma^0$ and $H_r\equiv
\sqrt[]{\mathstrut 8\pi G \rho_{\rm rad}^0/3}$, where the superscript 0 denotes present
values ($z=0$), therefore $n_\gamma^0$ and
$\rho_{\rm rad}^0$ are
 the present CBR photon number density and present radiation energy
 density, respectively.  Then
$n_X=n_X^0(1+z)^3\exp(-t/\tau_X)$ is transformed into $n_X=n_X^0(1/(2H_r
t))^{3/2}\exp(-t/\tau_X)$, so that the rate equation \ (\ref{dydt}) becomes
\begin{equation}
\frac{d Y_A}{d t} = \sum_P N_A(P)\left(-\frac{Y_A}{N_A(P) !} 
\left[A\gamma\right]_P + \frac{Y_P}{N_P(P) !}\left[P\gamma\right]_A\right),
\label{dydt2}
\end{equation}
where we have defined the reaction rate
\begin{eqnarray}
\left[B\gamma\right]_C&
\hspace{-5pt}
=&
\hspace{-5pt}
\frac{n_\gamma^0 \zeta_X}{\tau_X}
 \left(\frac{1}{2H_r t}\right)^{3/2} \exp(-t/\tau_X)
\nonumber \\
&&
\hspace{-5pt}
\times \int_{0}^\infty dE_\gamma
\left(\frac{\tau_X}{E_{\gamma 0}n_X}{\mathcal
N}_\gamma^{\rm QSE}(E_\gamma)\right)\,  \sigma_{\gamma +B \rightarrow C}
(E_\gamma),
\nonumber \\
\label{agamma}
\end{eqnarray}
where $\zeta_X=rE_{\gamma0}$.

The primary reactions and their cross sections are taken from~\cite{Cyburt:2002uv}.

\subsection{Secondary nucleosynthesis}
If the photo-dissociated light nucleus of a primary reaction has enough energy to
induce further nuclear reactions, then secondary or tertiary processes are possible.

The equation describing the secondary production and destruction is
obtained by taking into account the energy
loss of nuclear species while propagating through the background.  In
most situations, the energy loss rate is faster than any sink of
primary nuclei~\cite{Cyburt:2002uv} so that they can be ignored in the evolution of
the primary particles.  However, for unstable particles, the sink terms
must be evaluated.  In general,
because of the high reaction rate, the primary particles
establish a quasi-static equilibrium.  The abundance evolution is then represented by
\begin{equation}
\frac{d Y_S}{d t} = \frac{N_A(A)}{N_T(A) !} \frac{N_S(S)}{N_A(S) !}
 \sum_{T,T'}Y_T Y_{T'}[TT']_S -({\rm sink~term}),
\label{dysdt}
\end{equation}
where $N_K(L)$ has the same meaning as above. The quantity $[TT']_S$ is given by,
\begin{eqnarray}
\left[TT'\right]_S 
\hspace{-5pt}
&=&
\hspace{-5pt}
 \frac{\eta (n_\gamma^0)^2 \zeta_X}{\tau_X}
 \left(\frac{1}{2H_r t}\right)^3 \exp(-t/\tau_X)
\nonumber \\
&&
\hspace{-5pt}
\times \int_{0}^\infty dE_A \frac{\sigma_{A+T'\rightarrow S}(E_A)\beta_A}
 {b_A(E_A)} 
 \nonumber \\ 
&&
\hspace{-5pt}
 \times
  \int_{{\mathcal E}_A^{-1}(E_A)}^\infty \!\!\!\!\! dE_\gamma
  \left(\frac{\tau_X}{E_{\gamma 0}n_X}{\mathcal N}_\gamma^{\rm QSE}
   (E_\gamma)
\hspace{-3pt}
\right)
\hspace{-3pt}
 \sigma_{\gamma + T\rightarrow A}
  (E_\gamma)
\nonumber \\
&&
\hspace{-5pt}
\times \exp{\left[
	       -\int_{E_A}^{{\mathcal E}_A(E_\gamma)} 
	       dE_A^{^{\prime \prime}}
	       \frac{\Gamma_A(E_A^{^{\prime \prime}})}
	       {b_A(E_A^{^{\prime \prime}})}
	      \right] }~~.
  \label{tt'}
\end{eqnarray}
This is the reaction
rate for a secondary process $T(\gamma,X_1)A(T',X_2)S$ with any combination of 
particles $X_1$, $A$, and $X_2$.  The quantity $\eta$ is the 
baryon-to photon ratio:~$\eta\equiv n_B^0/n_\gamma^0$, and $\beta_A$
is the velocity of the primary particle $A$.  $b_A=-dE/dt$ is the rate of
energy loss of the primary particle.  The energy loss rates are taken
from~\cite{Kawasaki:2004qu}.  Coulomb
scattering ($N e^\pm_{\rm bg}\rightarrow N e^\pm$) is the dominant loss
process for primary nuclei $N$.  Then $\Gamma_A$ is the coefficient of the
primary particle sink, and in the case of an unstable nuclide, its value
is the decay rate.  ${\mathcal E}_A(E_\gamma)$ is the
energy of the nuclide $A$ produced by the photo-dissociation process
$\gamma+T\rightarrow A$, simultaneously ${\mathcal E}_A^{-1}(E_A)$ is
the energy of the non-thermal photons which produce the primary species
$A$ with energy $E_A$.

To simplify Eq.\ (\ref{tt'}) we define
\begin{eqnarray}
S_\gamma^{\rm QSE}(E_\gamma)=\frac{\tau_X}{E_{\gamma 0}n_X}{\mathcal N}_\gamma^{\rm QSE}(E_\gamma),
 \label{sgamma}
\end{eqnarray}
and
\begin{eqnarray}
 P(E_A,E_\gamma)=\exp{\left[
		       -\int_{E_A}^{{\mathcal E}_A(E_\gamma)} 
		       dE_A^{^{\prime \prime}}
		       \frac{\Gamma_A(E_A^{^{\prime \prime}})}
		       {b_A(E_A^{^{\prime \prime}})}
		      \right] }.
  \label{p_decay}
\end{eqnarray}
Then we denote the factors preceding the first integral
by an overall normalization $B$ to obtain
\begin{eqnarray}
\left[TT'\right]_S
\hspace{-5pt}
&=&
\hspace{-5pt}
B\int_{E_{p,\rm th}}^{{\mathcal E}_A(E_C)}\!\!\!\!\! dE_A 
 \frac{\sigma_{A+T'\rightarrow S}(E_A)\beta_A}{b_A(E_A)}
\nonumber \\
&&
\hspace{-5pt}
\times \int_{{\mathcal E}_A^{-1}(E_A)}^{E_C} \!\!\!\!\! dE_\gamma
 S_\gamma^{\rm QSE}(E_\gamma)\sigma_{\gamma + T\rightarrow A}
 (E_\gamma) P(E_A,E_\gamma),
 \nonumber \\
 \label{tt'2}
\end{eqnarray}
where lower
limit for the external integral is replaced with the threshold energy of
primary nuclide for the secondary reaction $E_{p,\rm th}$.  Since the photon spectrum includes few
energetic photons above $E_C$, the upper limit for
internal integral can be changed to $E_C$.  ${\mathcal E}_A(E_\gamma)$
and ${\mathcal E}_A^{-1}(E_A)$ are derivable in the limit of low energy
scattering, where the relevant nuclei are non-relativistic.  The
adoption of this limit is reasonable because most of the reactions occur
with low energy photons near threshold.  Thus, the
produced primary nuclei are mostly non-relativistic.

If the primary product particles are further destroyed by interactions with
background particles, this secondary destruction process can have a large contribution to
non-thermal nucleosynthesis.  Thus we also take into account the
destruction of D, T,$^3$He and $^6$Li after primary production by abundant
background nuclides.  For the destruction processes d+p and $^3$He+p, we used the cross sections
from~\cite{1972A&AS7417M,1974ARA&A12437R}.  For
 the t(p,dp)n and t(p,p2n)p reactions, we used cross sections from the mirror nucleus
reactions $^3$He(p,dp)p and $^3$He(p,2pn)p, respectively.  Although the cross
sections of t+p and $^3$He+p reactions are different because of the
Coulomb interaction, these mirror nucleus reaction
cross sections are almost identical~\cite{2002ApJ57689F}.  We adopt the cross section for
 $^6$Li(p,$^3$He)$^4$He
from~\cite{abr84}.

It has been found that only $^6$Li can be produced by secondary
non-thermal nucleosynthesis at a much higher level than by the SBBN,
while the production of other nuclei e.g. D, $^7$Li and $^7$Be are
insignificant~\cite{Jedamzik:1999di,Cyburt:2002uv,Kawasaki:2004qu}.  The relevant processes in
the secondary non-thermal production of $^6$Li involve interactions of
background $^4$He with primary tritium or $^3$He particles.  Threshold
energies of the $^4$He(t,n)$^6$Li and
$^4$He($^3$He,p)$^6$Li reactions are 8.3870 MeV and 7.0477 MeV,
respectively.  We have taken into account these
two reactions with their cross sections from~\cite{Cyburt:2002uv}.  

The survival
probability should also be considered in order to calculate the precise
abundance~\cite{Kawasaki:2004qu}.  This destruction of $^6$Li is included in the SBBN
code~\cite{Kawasaki:2004qu}.  Therefore the formulation of the survival probability should not
include the process after the thermalization of $^6$Li to prevent a
double counting of $^6$Li destruction.

\section{Observations of Light Element Abundances}\label{sec3}
\subsection{Light element abundances}\label{constraints}
The primordial abundances of D, $^3$He, $^4$He, and $^7$Li are inferred from
various observations.  Recently $^6$Li has been measured for numerous old
halo stars and a
so-called $^6$Li plateau as a function of metallicity appears to
exist~\cite{Asplund:2005yt}.  The reliability of inferred primordial abundances from several types of observations is
difficult to estimate because of systematic errors.  Here, we summarize the
observational data and our adopted constraints.

Deuterium is measured in absorption spectra at high redshift toward
QSOs.  Since the ratio D/H is
low, D can only be detected in absorption systems with high
HI column densities.  This causes a difficulty for such
observations.  From the Keck I HIRES spectra of Q1243+3047, a deuterium
abundance of D/H$=2.42^{+0.35}_{-0.25}\times 10^{-5}$ is
estimated~\cite{Kirkman:2003uv}.  The best estimate of the primordial D/H
from absorption systems toward five QSOs
reported in that paper, is
\begin{eqnarray}
 {\rm D}/{\rm H}=2.78^{+0.44}_{-0.38}\times 10^{-5}~~.
 \label{deuterium}
\end{eqnarray}
However, the five values have a larger dispersion than that inferred
from the individual measurements.  Hence, the errors are probably
underestimated.  We therefore take the highest
value of D/H in the five values, D/H$=3.98^{+0.59}_{-0.67}\times10^{-5}$
and allow for a two sigma increase above this value to fix an upper
limit.  In the
standard theory of galactic chemical evolution, deuterium continually
decreases in time by the processing of interstellar material
through stars.  Hence, we take the present abundance of deuterium as the
lower limit.  Column densities of DI along seven lines of sight
have been estimated from observations with the Far Ultraviolet
Spectroscopic Explorer (FUSE), and the local interstellar medium at a
distance of 37-179~pc has been probed~\cite{Moos:2001fe}.  The weighted mean value
of DI/HI for five data, with reliable values for column
densities of HI determined, is $(1.52\pm 0.08)\times 10^{-5}$.  When we
allow for a two sigma lower limit, we get the following constraint on the primordial abundance
of deuterium
\begin{eqnarray}
 1.4\times 10^{-5}< {\rm D}/{\rm H}< 5.2\times 10^{-5}~~.
 \label{deuterium2}
\end{eqnarray}

$^3$He is measured in Galactic HII regions by the 8.665~GHz
(3.46~cm) hyperfine transition of $^3$He$^+$~\cite{ban02}.  A plateau
with a relatively large dispersion with
respect to metallicity has been found at a level of $^3$He/H=$(1.9\pm 0.6)\times
10^{-5}$.  There is a problem, however.   Although stars are
thought to produce $^3$He, and $^3$He enhancement is observed in planetary
nebulae, the chemical evolution of $^3$He has not been detected in the Galaxy
during the last 4.5~Gyr~\cite{gei98,glo98}.  This fact has recently been
confirmed by more precise determination of the helium isotopic
composition of the local interstellar cloud~\cite{bus06}.  It is not yet
understood, therefore, whether $^3$He has increased or decreased through the course
of stellar and galactic chemical
evolution~\cite{Chiappini:2002hd,Vangioni-Flam:2002sa}.  Thus, we adopt the two sigma upper
limit from Galactic HII region abundances, that is
\begin{eqnarray}
 ^3{\rm He}/{\rm H}< 3.1\times 10^{-5}.
 \label{he3}
\end{eqnarray}
We will discuss later an implication for a tighter constraint as suggested
in~\cite{bus06}, i.e. $^3$He/H$<(1.6\pm0.3)\times 10^{-5}$, too.  We do not give a lower limit due to 
the large uncertainty in the galactic production of $^3$He.

$^4$He is measured in the HII regions of metal-poor external galaxies
where chemical evolution is thought to be minimal.  The primordial
abundance is estimated to be $Y=0.2421\pm0.0021$ by extrapolating the
abundance to zero metallicity (O/H=0)~\cite{Izotov:2003xn}.  However, in \cite{oli04}
it is noted that there are sources of systematic uncertainty in
determinations of the $^4$He abundance.  They suggest somewhat
 larger error bars with an abundance of
$Y=0.249\pm 0.009$.  They thus adopt a
primordial $^4$He abundance within the conservative range of
\begin{eqnarray}
 0.232< Y< 0.258.
 \label{he4}
\end{eqnarray}
We also adopt this constraint for the primordial $^4$He abundance.

$^7$Li is measured in metal-poor halo stars (MPHSs) by the spectra of their
atmosphere.  There is about a factor of three under-abundance of $^7$Li in
MPHSs with respect to the SBBN prediction when using the
baryon-to-photon ratio $\eta$ inferred from the analysis of the CMB
anisotropy.  This is called the lithium
problem~\cite{Ryan:1999vr,Melendez:2004ni,Asplund:2005yt}.  For a recent review of the lithium
problem, see~\cite{Lambert:2004kn}, where possible resolutions of this
problem are discussed.  

Recently, high
quality spectra for 24 metal-poor halo dwarfs and sub-giants have been
obtained~\cite{Asplund:2005yt}.  They estimated a mean $^7$Li
abundance of $\log \epsilon_{^7{\rm Li}}=2.21\pm 0.07$.  Different
groups derive somewhat different values.  We adopt the
estimate of \cite{Asplund:2005yt} for the primordial $^7$Li abundance 
allowing for a two
sigma range, $\log \epsilon_{^7{\rm Li}}=2.07-2.35$ or
\begin{eqnarray}
 1.17\times 10^{-10}<^7{\rm Li}/{\rm H}< 2.23\times 10^{-10}.
 \label{li7}
\end{eqnarray}
In addition, we take into account the possibility of the stellar depletion of
lithium up to 0.5~dex to derive an upper limit to the primordial abundance,
\begin{eqnarray}
 1.1\times 10^{-10}<^7{\rm Li}/{\rm H}< 7.1\times 10^{-10}.
 \label{li72}
\end{eqnarray}

As for the depletion of lithium in halo stars, it has been reported~\cite{cha05} that
the mean lithium abundance and its dispersion appear to be lower for 
dwarf stars than for turn-off and sub-giants.  $^7$Li
abundances of 28 such halo sub-giants have been measured~\cite{yon05}.  The result is that 
(excluding the extremely lithium-rich sub-giant BD~+23~3912) the
maximum abundance is $\log \epsilon_{^7{\rm Li}}=2.35$.  This is
well below the SBBN predicted value.  A calculation of the stellar
depletion of lithium isotopes by atomic and
turbulent diffusion leads~\cite{Richard:2004pj} to an abundance reduction by a factor of at
least 1.6-2.0 of the $^7$Li abundance for Population II stars with
metallicity [Fe/H]~$\leq -1.5$.

$^6$Li has also been measured in MPHSs by spectroscopy.  In~\cite{Asplund:2005yt}, $^6$Li
was detected at a better than two sigma significance in nine of the 24
stars observed.  They suggest that a $^6$Li plateau exists at $\log
\epsilon_{^6{\rm Li}}\approx 0.8$.  This plateau implies important
information because the SBBN would predict that much less primordial $^6$Li be formed
($^6$Li/$^7$Li $\sim 10^{-5}$).  Therefore,
some mechanism should have produced almost all $^6$Li in MPHSs.  There
are several candidates.  For example, gravitational
shocks during the hierarchical structure formation of the Galaxy
could have accelerated cosmic rays (CRs) and produced $^6$Li by
$\alpha-\alpha$ fusion reactions at a
very early stage~\cite{suz02,ino05}.  A scenario has also been
suggested~\cite{Rollinde:2004kz,Rollinde:2006zx} whereby cosmological CRs (perhaps related to
pop III stars) were produced at an extremely early epoch
up to the formation of the Galaxy.  These CRs could then have produced $^6$Li by $\alpha+\alpha$
fusion.  See the discussion
in~\cite{Prantzos:2005mh} for a recent summary of various candidates for $^6$Li
production.  Another new mechanism has been proposed recently~\cite{Pospelov:2006sc} whereby $^6$Li originated from the binding of negatively
charged particles to background nuclei.  Since multiple processes have possibly synthesized $^6$Li at
an early
epoch, we do not put limits on the primordial abundance of
$^6$Li.  However, we adopt the average value of the abundance derived from
the eight MPHSs with detections as a guide to constrain possible production from radiative decay,
\begin{eqnarray}
 ^6{\rm Li}/{\rm H}\approx 6.6\times 10^{-12}.
 \label{li6}
\end{eqnarray}

\subsection{Cosmic microwave background anisotropy}
Very precise data have been obtained by observations of the spectrum of temperature fluctuations in the CMB.  The WMAP data have been 
analyzed and the energy density of baryons in the
universe have been deduced along with other cosmological parameters.  This leads to $\Omega_b h^2=0.0224\pm 0.0009$ for the WMAP first year
data~\cite{Spergel:2003cb} and $\Omega_b h^2=0.0207^{+0.0008}_{-0.0011}$
for the WMAP three year data~\cite{Spergel:2006hy} in the running scalar
spectral index model.  We adopt the
first year result, corresponding to
$\eta=(6.1^{+0.3}_{-0.2})\times 10^{-10}$.  The SBBN with the
WMAP $\Omega_b h^2$ parameter region has been calculated including the uncertainties of
the inferred $\Omega_b h^2$ and the the reaction
rate uncertainties on the SBBN~\cite{Coc:2003ce}.  Their result is:
\begin{eqnarray}
 {\rm D}/{\rm H}&=& \left(2.60^{+0.19}_{-0.17}\right)\times 10^{-5}
~\label{wmap1}
\\
 ^3{\rm He}/{\rm H}&=& \left(1.04\pm 0.04\right)\times 10^{-5}
~\label{wmap2}
\\
 Y&=& 0.2479\pm 0.0004
~\label{wmap3}
\\
 ^7{\rm Li}/{\rm H}&=& \left(4.15^{+0.49}_{-0.45}\right)\times 10^{-10}.
~\label{wmap4}
\end{eqnarray}

\section{Results}\label{sec4}
We have calculated the cosmological nucleosynthesis including the SBBN and
non-thermal nucleosynthesis induced by the radiative decay of a
long-lived massive particle.  The SBBN was computed using the 
Kawano code~\cite{kawano} with the use of the new world average of the
neutron lifetime~\cite{Mathews:2004kc}.  We added the non-thermal components of Eq.\
(\ref{dydt2}) and (\ref{dysdt}) also taking into account the destruction of
secondarily produced $^6$Li.  We deduced properties of the decaying particle in terms of its lifetime
$\tau_X$ and abundance $\zeta_X$ with the baryon-to-photon ratio fixed at
$\eta=6.1\times 10^{-10}$.  We checked the effect of secondary
destruction of the primary non-thermal nuclides.  We confirmed that the energy
loss is much faster than the destruction (see the discussion for
non-thermal $^6$Li production of~\cite{Cyburt:2002uv}).  The secondary
destruction processes of primary nuclides were not very efficient (destruction probabilities are
$\leq {\cal O}(10^{-3})$).  The final abundances were therefore not much
affected by secondary destruction.

The point is that the time scale of the Coulomb
loss for the non-thermal nuclides is much smaller than those of the destruction reactions at low
energies.  Furthermore, the
destruction probabilities of non-thermally produced $^6$Li are very
small, and produced $^6$Li survives~\cite{Kawasaki:2004qu}.

We have derived the constraints on the lifetime $\tau_X$ and abundance parameter
$\zeta_X$ from the adopted limits for the cosmological light element abundances.  
Our result is very similar to that
of~\cite{Cyburt:2002uv}, since we use the same formulation for non-thermal
nucleosynthesis and adopt their fitted cross sections.  A detailed explanation has
been given in~\cite{Cyburt:2002uv} for the
systematics of the radiative decay.

\begin{figure}[tbp]
\rotatebox{-90}{\includegraphics[height=8.0cm,clip]{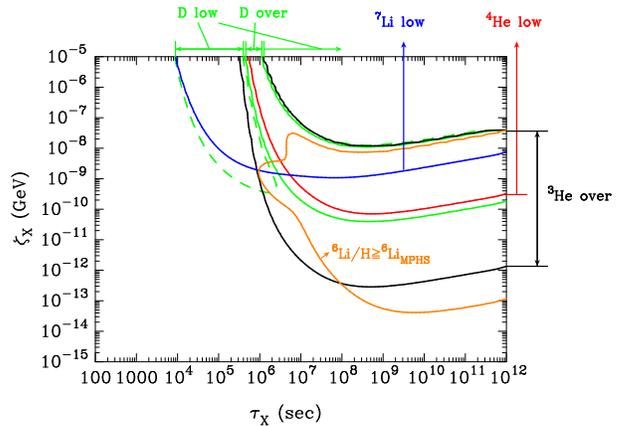}}
\caption{\label{contour} Contours in the $(\tau_X,\zeta_X)$
 plane corresponding to the adopted constraints for the primordial
 abundances in models with a fixed value of $\eta=6.1\times 10^{-10}$.  Contours for the mass
 fraction of $^4$He $Y=0.232$ (red line) and the
 number ratios of $^3$He/H=$3.1\times 10^{-5}$ (black lines),
 D/H=$5.2\times 10^{-5}$ (green solid lines), D/H=$1.4\times 10^{-5}$ (green dashed lines), and $^7$Li/H=$1.1\times 10^{-10}$ (blue line) are
 shown.  The contour of $^6$Li/H=$6.6\times 10^{-12}$ (orange line) is
 also drawn.  The notation ``over'' and ``low'' identifies overproduced
 and underproduced regions, respectively.}
\end{figure}

Fig.~\ref{contour} shows a contour of the $^4$He mass fraction
$Y>0.232$ (red line) in the
($\tau_X$,$\zeta_X$) plane.  Above this contour, $Y<0.232$.  Since $^7$Li
is more weakly bound than $^4$He,
the destruction happens even in shorter lifetime conditions.  The
contour of $^7$Li for the lower limit of $^7$Li/H$>1.1\times 10^{-10}$
(blue line)
has the shape plotted on Fig.~\ref{contour}.  Contours for D/H upper and
lower limits, D/H$ \leq5.2\times
10^{-5}$ (green solid lines) and D/H$ \geq1.4\times 10^{-5}$ (green dashed lines),
respectively are also shown.  Contours for the $^3$He/H upper limit,
$^3$He/H$ \leq 3.1\times
10^{-5}$ (black lines) are also drawn.  The shape of this contour is very analogous to that
for the D/H upper limit.  This reflects the fact that the photodisintegration of
$^4$He is the main cause of the production of both D and $^3$He.  Finally, the contour
for the MPHSs level of $^6$Li/H=$6.6\times 10^{-12}$ is plotted to see the behavior of the $^6$Li non-thermal production 
(orange line).

\begin{figure}[tbp]
\rotatebox{-90}{\includegraphics[height=8.0cm,clip]{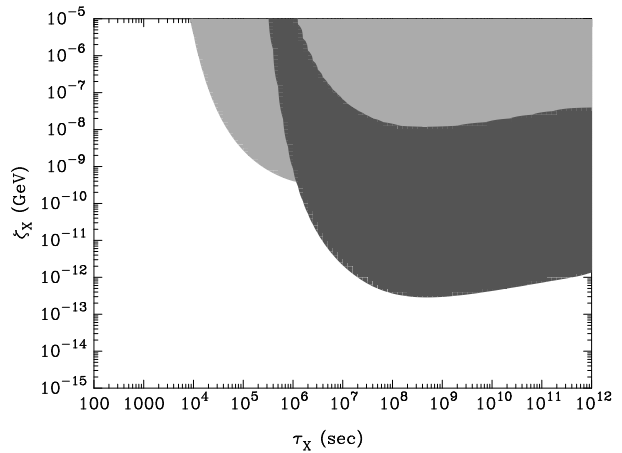}}
\caption{\label{cons_nucl} Colored regions identify the 
excluded area in the parameter space ($\tau_X$,$\zeta_X$) for models with
a fixed value of $\eta=6.1 \times 10^{-10}$.  The dark shaded
 region is excluded by an overabundance of $^3$He, whereas the light
 shaded region is mostly excluded by an underabundance of
 deuterium.}
\end{figure}

Fig.~\ref{cons_nucl} shows the derived constraint on $\tau_X$ and
$\zeta_X$ for an unstable particle from the consideration of the light
element abundances above described in a model with $\eta=6.1 \times
10^{-10}$.  All limits from the light element
abundances used in this study, for D/H, $^3$He/H, $Y$ and $^7$Li/H
have been taken into account.  The $^3$He overabundant region is shaded by the dark
color, and the rest of the excluded region a light color.  The light colored
region is fixed largely by the deuterium underproduction.  For
$\tau_X$$^>_\sim 10^{6}$~s, $^3$He provides the strongest limit on the
abundance parameter yielding, 
\begin{eqnarray}
 \zeta_X\leq 4\times10^{-13}~{\rm GeV}~~,
 \label{cons5}
\end{eqnarray}
while for shorter lifetimes ($\tau_X\sim 10^4-10^6$~s) the limits
are from D implying $\zeta_X~  ^<_\sim  10^{-9}~{\rm GeV}$.

Uncertainties
from the reaction cross sections have been discussed
in~\cite{Cyburt:2002uv}.  The
uncertainties from the photodissociation and secondary cross sections
are of the order of 10~\%.  In addition, we confirmed that effects of errors in the cross
sections for the secondary destruction reactions are negligible.  The BBN
 light element abundances are then the dominant source of
uncertainty.  The BBN cross
section uncertainties are not included in our study, but they are
estimated [e.g. Eq.\ (\ref{wmap1})-\ (\ref{wmap4})], and one can apply
those to our non-thermal
calculation results.

\section{Discussion}\label{sec5}

\subsection{Distortion of the CMB spectrum}
The abundance of massive decaying particles is also constrained
from the observed spectrum of CMB radiation.  Since non-thermal photons produced by the
radiative decay deform the blackbody spectrum of the CMB, this is
limited by the consistency of the observed CMB data with a blackbody
spectrum~\cite{Hu:1993gc,Feng:2003xh}.  For epochs earlier than $z\sim10^7$,
thermal bremsstrahlung, [i.e. free-free emission ($eN\rightarrow eN\gamma$),
where $N$ is an ion] and radiative-Compton scattering ($e^-\gamma\rightarrow e^-\gamma\gamma$) act effectively to erase any
distortion of the CBR spectrum from a blackbody.  For the decay in
epochs $10^5<z<10^7$, processes changing the photon number become ineffective,
and Compton scattering ($\gamma e^-\rightarrow \gamma e^-$) causes
the photons and electrons to achieve statistical equilibrium, but not
thermodynamic equilibrium.  Then, the photon spectrum should have a
Bose-Einstein distribution
\begin{eqnarray}
 f_\gamma(\vec{p}_\gamma)=\frac{1}{e^{\epsilon_\gamma/T+\mu}-1}~~,
 \label{distortion1}
\end{eqnarray}
where $\mu$ is the usual dimensionless chemical potential derived from the
conservation of photon number.

Analyses of the CMB data suggest a relatively low baryon density so that
double Compton scattering dominates the thermalization process.  For 
small energy injection from the radiative decay, the chemical potential
can be
approximated analytically~\cite{Hu:1993gc,Hu:1992dc} as
\begin{equation}
 \mu=4.0\times 10^{-4}\left[\frac{\tau_X}{10^6~{\rm s}}\right]^{1/2}
  \left[\frac{\zeta_X}{10^{-9}~{\rm GeV}}\right]e^{-(\tau_{\rm dC}/\tau_X)^{5/4}}~~,
 \label{distortion2}
\end{equation}
where
\begin{equation}
 \tau_{\rm dC}=6.1\times10^6~{\rm s} 
\hspace{-3pt}
\left[\frac{T_0}{2.725~{\rm K}}\right]^{
\hspace{-2pt}
-12/5}
\hspace{-3pt}
 \left[\frac{\Omega_b h^2}{0.022}\right]^{
\hspace{-2pt}
4/5}
\hspace{-3pt}
 \left[\frac{1-Y/2}{0.88}\right]^{
\hspace{-2pt}
4/5}~~,
 \label{distortion3}
\end{equation}
where $T_0$ is the present CMB temperature, and $h=H_0/(100~{\rm km~s}^{-1}~{\rm
Mpc}^{-1})$ is the normalized Hubble parameter.

For a late energy injection at $z<10^5$, Compton scattering produces
little effect and cannot establish a Bose-Einstein spectrum.  The
distorted spectrum is then described by the Compton $y$ parameter.  There
is a relation between $y$ and the amount of the injected energy, $\Delta E/E_{\rm CBR}=4y$, where $\Delta E$ and $E_{\rm CBR}$ are the
total energy injected and the CBR energy, respectively.  The ratio of the energy injected by radiative decay to the CBR energy
per comoving volume can be expressed as
\begin{eqnarray}
 \frac{\Delta E}{E_{\rm CBR}}=\frac{E_{\gamma0}}{2.7T(t_{\rm eff})}
  \left[\frac{n_{X0}}{n_{\gamma0}}\right]=\frac{\zeta_X}{2.7T(t_{\rm eff})}~~,
 \label{distortion4}
\end{eqnarray}
where $t_{\rm eff}=[\Gamma(1+\beta)]^{1/\beta} \tau_X$ for
$T\propto t^{-\beta}$, with $\Gamma(x)$ the gamma function of
argument $x$.  The factor of 2.7 comes from the fact that the average energy
of the CBR photons at a given temperature is $2.7T$.  The time-temperature relation is $T\propto t^{-1/2}$ for a radiation dominated universe.

\begin{figure}[tbp]
\rotatebox{-90}{\includegraphics[height=8.0cm,clip]{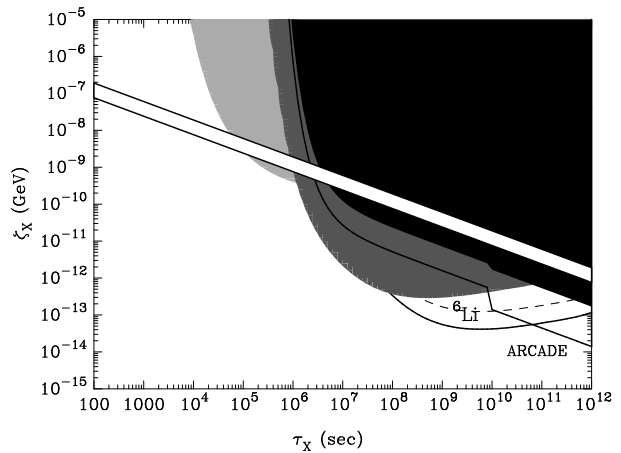}}
\caption{\label{cons_cmb} Same as in Fig.~\ref{cons_nucl} except that the black shaded
 region is superimposed.  This region shows the region excluded by the consistency requirement 
 of the CMB with a
 blackbody.  The white band identifies parameter values where the free-streaming of 
 decay products would lead to a suppression of small
 scale structure growth in a model with $\Omega_Y h^2\sim 0.11$.  The curved line identifies the contour of $^6{\rm
Li}/{\rm H}=6.6\times 10^{-12}$, corresponding to the abundance of $^6$Li observed
in MPHSs.  The
region above the contour and below the nucleosynthesis and CMB
constraints is allowed and abundant in $^6$Li.  The contour for $^6$Li enhanced by three
 times as much MPHSs value is also drawn (dashed line).  The solid line
corresponds to the sensitivity of the planned ARCADE mission as labeled.}
\end{figure}

The CMB spectrum has been well measured and the deduced limits are $|\mu|<9\times 10^{-5}$, $|y|<1.2\times
10^{-5}$~\cite{Hagiwara:2002fs} and $\Omega_b h^2\sim 0.022$ with $h\sim
0.71$~\cite{Spergel:2003cb}.  Therefore, the high abundance parameter region of
$\zeta_X$ is excluded by the $\mu$ and $y$ limits.  In Fig.~\ref{cons_cmb} the black shading indicates 
the parameter region excluded by the CBR distortion limit.  For a lifetime shorter than $\tau_X=4\times 10^{11}~{\rm s}~\Omega_b h^2\sim
8.8\times 10^9~{\rm s}$, the decay is constrained by the
chemical potential $\mu$.  On the other hand, when an unstable particle
decays later, the CBR spectrum is limited by the Compton $y$ parameter.
The parameter
region of relatively long lifetime ($10^{10}~{\rm s}<\tau_X$) is found
to be constrained by the CMB spectrum more strongly than the light
element abundances.

The discussion so far has assumed that $\eta$ at the BBN epoch can be
determined from the power spectrum of CMB
temperature fluctuations.  However, $\eta$ could have changed from the BBN epoch to the time of the
recombination~\cite{Feng:2003uy}.  The entropy production
resulting from the photon emission by radiative decay has
been estimated and it is very small for a large part of the
region allowed by the non-thermal nucleosynthesis and CMB distortion
constraints in Fig.~\ref{cons_cmb}.  However, for very early decay
($\tau_X < 10^4$~s) entropy production could be large.  An entropy change
by as large as $\Delta S/S_i\sim 10^{-1}$ seems unlikely since the
CMB-favored value of $\eta=6.1 \times 10^{-10}$ is transformed into
$6.7 \times 10^{-10}$ at the BBN epoch and this would bring about a more severe
discrepancy of the $^7$Li abundances between the SBBN
prediction including the non-thermal processes and estimations from the
MPHSs observations.  See Fig.~3 in~\cite{Feng:2003uy}

\subsection{Effects on small-scale structure}
The decay products obtain momentum
from the decay by simple momentum conservation.  This momentum
can affect structure formation in the universe.  In this regard there are useful constraints on the
present root-mean-square velocity of warm dark matter from observations of the
 Lyman-$\alpha$ forest~\cite{Viel:2005qj} and super-massive black
holes at high redshift~\cite{Barkana:2001gr}.  Limits from the early reionization
have also been reported~\cite{Barkana:2001gr,Yoshida:2003rm}.  The free-streaming 
decay products must not cause an inconsistency with observations.  This
consideration of free-streaming has been
discussed in detail in~\cite{Jedamzik:2005dh}.

Cold, collision-less, non-self-interacting dark matter explains the
trend of the observed structures larger than $\sim 1$~Mpc.  However, it
does not successfully explain the structure at smaller scales.  As a solution to the small
scale structure problem, the decay of SWIMPs has been
proposed~\cite{Cembranos:2005us}.  The velocity dispersion reduces the
6-dimensional phase space density and prevents the formations of cuspy
halos.  Furthermore, the free-streaming of dark matter particles damps the
power spectrum at small scales.  They discuss the paradigm in which the dark matter is
composed mostly of the decay products.  In this case the parameters
which lead to reasonable free-streaming predict roughly the correct phase
space density.  Therefore we only discuss the free-streaming scale.

If the free-streaming scale is reasonably small $\sim 1$~Mpc, the
free-streaming scale $ \lambda_{\rm FS}$ from the decay time to the
present ($z\sim0$) is described~\cite{Cembranos:2005us} as
\begin{eqnarray}
 \lambda_{\rm FS}\approx1.5~{\rm Mpc}~u_Y \left[\frac{\tau_X}{10^6~{\rm s}}\right]^{1/2}~~,
 \label{free}
\end{eqnarray}
where $u_Y\equiv p_Y/m$ is momentum at the decay time divided by the
mass of the particle $Y$ produced by the decay.  In the range of
$\tau_X\leq 10^{12}$~s, this approximation involves a maximum error of only a factor 1.5 in
the numerical factor.  The largest error comes when $\tau_X=10^{12}$~s
and correspondingly $u_Y\sim 10^{-3}$.  From momentum conservation during
the decay, the momentum of product $p_Y$ is equal to the emitted photon
energy $E_{\gamma 0}$.  The parameter $\zeta_X$ is written as
\begin{eqnarray}
 \zeta_X=2.6\times 10^{-8}~{\rm GeV}~(\Omega_Y h^2)u_Y~~,
 \label{free2}
\end{eqnarray}
where $\Omega_Y$ is the energy density normalized to the critical
density.  

Eqs.\ (\ref{free})
and\ (\ref{free2}) connect the $\tau_X$ and $\zeta_X$ by the elimination of
$u_Y$.  Such decay can be a
resolution to the small scale structure problems if the free-streaming
scale is 0.4~Mpc~$\alt \lambda_{\rm
FS}\alt$~1.0~Mpc~\cite{Cembranos:2005us}.  This limit determines the suitable
parameter space for $\tau_X$ and $\zeta_X$ under the assumption of
$\Omega_Y h^2\sim 0.11$~\cite{Spergel:2003cb}.  This
region is shown in Fig.~\ref{cons_cmb} as a white band bounded by
solid lines.  The region above this band is excluded in this case, since the decay
products erase the larger scale fluctuations.  The region below
this band will be discussed in the next subsection in connection with
the $^6$Li abundance.  Apparently, 
 the decay at $\tau_X\agt 10^5$~s is not
realistic because the effect on the light element abundances or CBR
spectrum is prohibitively large.  On the other hand relatively early
decay at $\tau_X\alt 10^5$~s is viable.  Applying this result to a
model for a gravitino SWIMP with a photino NLSP~\cite{Cembranos:2005us}, we obtain
$m_{\rm SWIMP}\alt 200$~GeV and $\Delta m\sim$~400-700~GeV.

\subsection{$^6$Li-producing parameter region}
In this paper we analyze the possibility that the radiative decay of long-lived
particles produces $^6$Li by non-thermal process while having almost no
effect on $^7$Li or other nuclides produced in the SBBN.  Ellis, Olive \&
Vangioni researched the possibility that the radiative decay of unstable particles explains the discrepancy of the BBN calculated $^7$Li abundance and low $^7$Li
plateau derived from observations~\cite{Ellis:2005ii}.  They found that 
in the parameter region where $^7$Li is
photo-dissociated down to the level of the $^7$Li plateau, either the D
abundance was too low or the ratio $^3$He/D was unacceptably
large in the context of standard stellar evolution and chemical
evolution.  Consequently, they concluded that radiative particle decays cannot be
a cause for the $^7$Li abundance difference.  They also mentioned the possibility of 
$^6$Li production in their paper.

Uncertainties remain in estimations of the Li abundance in stellar
atmospheres, and the probability of depletion in stars has not been
excluded.  The difference between abundances determined by different
analysis approaches is also a concern.  Therefore, we suppose that the discrepancy
of the $^7$Li abundance is caused by stellar depletion or some other systematic
effect.  Then the $^6$Li abundance in the early universe should have been
 larger when first engulfed in a
star than the value presently deduced from observations of
MPHSs.  Assuming that is the case, we impose the following constraint on the $^6$Li
abundance after the radiative decay process, 
\begin{eqnarray}
 ^6{\rm Li}/{\rm H}> 6.6\times 10^{-12}~~.
 \label{li_pro1}
\end{eqnarray}
If this process is responsible for the high $^6$Li abundance in MPHSs,
this is the limit to be adopted.  Although previous studies~\cite{Jedamzik:1999di,Ellis:2005ii} give the $^6$Li abundance of MPHSs as the
upper limit, we adopt this as the lower limit considering the possibility of stellar destruction. 

In Fig.~\ref{cons_cmb} the contour of the lower limit (\ref{li_pro1}) is
shown by a solid line below the CMB constraint.  Hence, a $^6$Li-producing
allowed parameter region certainly exists for $\tau_X=10^8-10^{12}$~s
and $\zeta_X\sim 10^{-13}-10^{-12}$~GeV.  The parameter region
 allowed by the above constraints which also produces abundant
$^6$Li is marked as ``$^6$Li''.

This $^6$Li-producing zone is somewhat close to the limit deduced by
fits to the CMB distortion.  Thus, future measurements of the CMB
spectrum may reach this parameter region and provide better 
limits to this radiative decaying particle scenario.  The Absolute Radiometer for Cosmology, Astrophysics, and
Diffuse Emission (ARCADE)~\cite{ARCADE,Kogut:2004hn} will observe the CMB at centimeter wavelengths with better
sensitivity than current data.  ARCADE may make it possible to impose a limit on the chemical
potential up to $|\mu|<2\times 10^{-5}$ and Compton $y$ parameter up to
$y< 10^{-6}$, respectively.  This sensitivity corresponds to the
solid line marked as ``ARCADE'' in Fig.~\ref{cons_cmb}.  It spans the
parameter region for producing three times as much $^6$Li as MPHSs (a dashed
line, see discussion below).  Thus, if
ARCADE does not detect the spectral distortion, $^6$Li production
from radiative decay can be ruled out.

The allowed parameter region has a very
small entropy increase of less than the order of $10^{-5}$, so that this
has a negligible effect on the deduced $\eta$.  We have analyzed this parameter region to see
the possibility of realization.  It is important to check the results of the non-thermal nucleosynthesis for
decay in the $^6$Li-producing region.  We made a model calculation with input parameters of $\tau_X=1\times
10^{10}$~s, $\zeta_X=3\times 10^{-13}$~GeV and $\eta=6.1\times
10^{-10}$.  The final
abundances obtained in this model are
\begin{eqnarray}
 {\rm D}/{\rm H}&=& 2.63\times 10^{-5}
\\
 ^3{\rm He}/{\rm H}&=& 2.48\times 10^{-5}
\\
 Y&=& 0.247
\\
 ^6{\rm Li}/{\rm H}&=& 4.69\times 10^{-11}
\\
 ^7{\rm Li}/{\rm H}&=& 4.36\times 10^{-10}.
 \label{li_pro2}
\end{eqnarray}

These are certainly consistent with the constraints we adopted in
Sec.~\ref{constraints}.  The abundances of $^3$He and $^6$Li with respect to the SBBN abundances
 increase.  In fact, the non-thermal production of $^6$Li is unavoidably
 accompanied by $^3$He production.  We confirmed that the produced amounts of
$^3$He and $^6$Li are proportional to $\zeta_X$ when $\tau_X$ is
fixed.  However, for a longer decay lifetime, the $^6$Li production is
relatively more effective than the $^3$He production because of
the different dependences of the production rates on the
CBR temperature.

If the inconsistency
between the $^7$Li abundance predicted by SBBN and that
measured from MPHSs is caused by depletion, $^6$Li
would have existed in the primordial gas at a level larger than the abundance observed in
MPHSs by at least the ratio of the
SBBN $^7$Li/H prediction to the mean value observed in MPHSs.  The observed $^7$Li/H 
abundance~\cite{Asplund:2005yt} is
 $^7$Li/H$\sim1.62\times
10^{-10}$.  Hence, this factor is about $(4.36\times 10^{-10})/(1.62\times
10^{-10})\sim 3$.  So $^6$Li should have been originally produced at an abundance more
than about
3 times the presently observed value.  We stress that the non-thermal $^6$Li production
inevitably brings about the production of $^3$He, and this gives a strong
constraint on the possible parameter space
 of unstable particles~\cite{Kawasaki:2004qu,Ellis:2005ii}.

\begin{figure}[tbp]
\rotatebox{-90}{\includegraphics[height=8.0cm,clip]{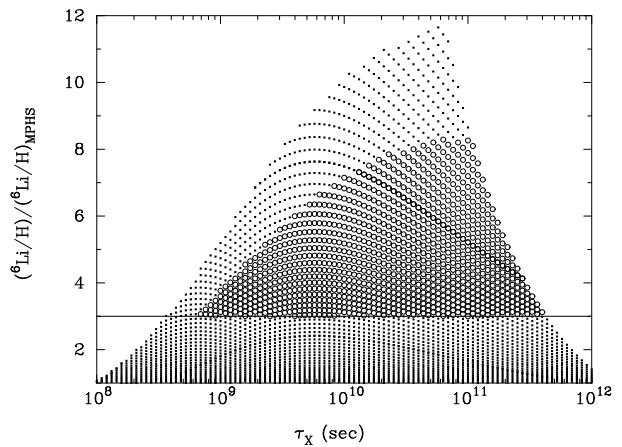}}
\caption{\label{relation2} Ratio of calculated $^6$Li/H abundances (after the non-thermal nucleosynthesis) to the observed abundance in MPHSs as a function of
 $\tau_X$.  Results in the allowed parameter
 region of ($\tau_X$, $\zeta_X$) producing $^6$Li/H larger than the value
 found in MPHSs, or the marked region ``$^6$Li'' in
 Fig.~\ref{cons_cmb} are plotted.  The horizontal line indicates a factor of three overproduction of
 $^6$Li relative to the observed MPHS value of $^6$Li/H$=3\times 6.6\times 10^{-12}$.  The
 large circles denote values in the allowed region with abundances of
 $^3$He/H=$1.3-2.5\times 10^{-5}$ and $^6$Li/H$\geq3\times 6.6\times 10^{-12}$.  The other
 parameters sets are indicated by small squares.}
\end{figure}

We have analyzed the upper limit to the $^6$Li abundance resulting from the
radiative decay process under the requirement of consistency with the other light-element
abundances.  In Fig.~\ref{relation2}, the $^6$Li abundances are plotted
as a function of $\tau_X$.  Points on this figure are allowed by the constraints
imposed above and lead to $^6$Li abundances above the level observed in
MPHSs.  The vertical scale is $^6$Li/H
normalized to the mean $^6$Li/H abundance in
MPHSs ($^6$Li/H)$_{\rm MPHS}$.  The horizontal line indicates
 a factor of three enhancement in $^6$Li.  The large circles
are for cases with more than three times as abundant $^6$Li as the level
found in MPHSs.  Here, we adopt the one sigma $^3$He/H=$(1.9\pm0.6)\times 10^{-5}$~\cite{ban02}
as an extra constraint.  We note that, in cases adopting a tighter
constraint $^3$He/H$<(1.6\pm0.3)\times 10^{-5}$~\cite{bus06}, one can
still find an allowed region of $\tau_X=3\times 10^{10}-3\times 10^{11}$~s
which satisfies the same constraint imposed on the $^6$Li abundance.  The small squares are for other cases of Eq.\
(\ref{he3}). 

This figure confirms that $^6$Li/H abundances as large as
those in MPHSs
multiplied by the ratio ($^7$Li/H)/($^7$Li/H)$_{\rm MPHS}$ can be produced
by non-thermal nucleosynthesis without significantly impacting the other nuclide
abundances.  Although this explanation could resolve the discrepancy between the
SBBN predicted $^6$Li abundances and those derived from observations, it
cannot resolve the lithium problem.  This scenario necessarily requires 
some model for the stellar depletion of $^6$Li and $^7$Li.  Indeed,
as discussed in~\cite{Asplund:2005yt} and references therein, models exist which suggest a very
large depletion factor of $^6$Li along with some $^7$Li
depletion.  The production of $^6$Li by radiative decay cannot explain the observed
abundances of both $^6$Li and $^7$Li, if the stellar depletion proceeds as
described by that model.  However, approximately equal amounts of depletion
for both lithium isotopes can explain the measured abundances when combined
with the non-thermal production of $^6$Li.  As for the case including the
hadronic decay process~\cite{Jedamzik:2005dh}, it
has been found that such particle decay could simultaneously solve both the $^6$Li and
$^7$Li problem, even if a possible degree of depletion is
included.  Clearly, further research on such depletion
processes, along with analysis of more observational data, is desired.

\section{Conclusion}\label{sec6}

We have investigated
the possibility that non-thermal nucleosynthesis induced by the
radiative decay of long-lived particles contributed to
 the light element abundances in the early universe.  We have constrained this process
 by observed primordial light element abundances. 
 The primordial nucleosynthesis induced by high energy
non-thermal photons from the radiative decay of long-lived relic
particles was calculated taking into account both the primary nuclear production reactions and the effects of secondary production and
destruction processes. 
We find, however, that the secondary destruction processes of primary D, T, $^3$He
and $^6$Li have little influence on the final light element
abundances and in particular, it is confirmed that the effect of the
destruction of the produced $^6$Li is very small in the cosmic
temperature regime where the non-thermal secondary $^6$Li production is operative.

We have utilized the observed light element abundances from
various sources to explore the constraints on the parameter space
of the lifetime ($\tau_X$) and initial abundance ($\zeta_X$) of unstable
particles in a model in which the baryon-to-photon ratio $\eta$ is fixed to the 
value inferred from the WMAP
CMB power spectrum.  We require that the non-thermal processes 
do not cause significant deviations
from the observationally inferred primordial abundances.  For each given
lifetime $\tau_X$ we deduce an upper limit to the product $\zeta_X$ 
of the photon energy of the radiative decay and the fraction
of the number density of the decaying particles to that of the CBR
photons.  For short lifetimes, $\tau_X<10^6$~s, the lower limit to $\zeta_X$
 is fixed by a D under-abundance, whereas for longer lifetimes,
$^3$He overproduction gives the strongest upper limit to $\zeta_X$.

The parameter values are also constrained by the induced distortion of the CMB
black-body spectrum from energetic photons emitted in the decay.  The CMB constraint is more important than the light element abundance constraints 
for the case of a long lifetime $\tau_X>10^{10}$~s.  Future missions such as
ARCADE would, perhaps, detect the signal of the radiative decay.  Otherwise,
the production of $^6$Li in this process is constrained. 

Next we
considered the free-streaming scale of the decay products.  We imposed a
 constraint on the abundance
parameter $\zeta_X$ under the
assumption that the decay products comprise a dominant constituent of
the cosmological dark matter.  For the specific model of
Ref.~\cite{Cembranos:2005us} a relatively late decay at $\tau_X>10^5$~s
is forbidden.

We studied the possibility of $^6$Li production in 
non-thermal nucleosynthesis at a level which is comparable with the observed abundance in
MPHSs.  We find that there exists a parameter region leading to final
abundances which are in reasonable agreement with
the MPHS observations of $^6$Li and is also consistent with the observational constraints on
the other light nuclides.  We show that if $\tau_X\sim
10^8-10^{12}$~s and $\zeta_X\sim 10^{-13}-10^{-12}$~GeV,
it is possible that the radiative decay of
long-lived relic particles causes the observed enhanced abundance of $^6$Li in
MPHSs. 

Analyzing the results of the light element abundances for the interesting
particle parameter region in more detail, two important characteristics were found:
One is that the excess $^6$Li abundances from the non-thermal processes
is regulated by the amount of $^3$He co-production because $^3$He is the
seed for $^6$Li in the process $^4$He($^3$He,p)$^6$Li.  Hence, the radiative
decay model which results in $^6$Li-production above the MPHS abundance
level is also reflected by an enhancement of the $^3$He abundance with respect to the SBBN
value.  Therefore, tighter constraints on the primordial $^3$He abundance
might exclude some parameter regions.  However, a certain range of the possible
parameter region still remains.

Another feature regarding this scenario of non-thermal $^6$Li
production triggered by the radiative decay is that it does not resolve the
lithium problem.  Other mechanisms such as the stellar depletion of the
lithium isotopes must operate to lower the $^7$Li abundance in the atmosphere of
MPHSs.  In such mechanisms, however, $^6$Li should be simultaneously
depleted.  In the parameter region,
$\tau_X\sim 10^8-10^{12}$~s and $\zeta_X\sim 10^{-13}-10^{-12}$~GeV, the
$^7$Li abundance is almost unchanged from the SBBN value even should the
radiative decay of unstable particles and subsequent non-thermal
nucleosynthesis be taken into consideration.  Only $^6$Li is
enhanced by about three orders of magnitude higher than the SBBN
value.  This enhancement can be as much as several times the observed $^6$Li
abundance in MPHSs.  Hence, the present scenario remains a viable possibility 
for $^6$Li enhancement if
the true depletion mechanisms or other systematic effects lead to
approximately the same degree of depletion for both $^6$Li and $^7$Li.

In summary, we have found a parameter region of $\tau_X\sim 10^8-10^{12}$~s
and $\zeta_X\sim 10^{-13}-10^{-12}$~GeV where the non-thermal
nucleosynthesis of $^6$Li can explain the observed abundance level in
MPHSs.  This parameter region satisfies the two observational
constraints on the CMB energy spectrum and the primordial light element
abundances, although it cannot be a solution to the small scale
structure problem.

It is very fascinating to consider the possibility that such new physics
as the decay of exotic particles plays a role in producing the observed
abundances of lithium in the universe.  Clearly, further research into lithium nucleosynthesis and the interpretation of the $^6$Li
plateau would be very valuable.  Also, further observations of the $^6$Li abundances in the
stellar atmosphere of MPHSs will help to constrain the various suggested candidates for $^6$Li
production and to answer the question as to whether it is synthesized
cosmologically or by cosmic ray interactions in the Galaxy. 
\begin{acknowledgments}
We are grateful to Veniamin S. Berezinsky for enlightening comments on
the electromagnetic cascade process, and to Michael A. Famiano for
providing information on the cross sections which we used in our
 study.  We thank David Yong for explaining to us his research on lithium
 observations.  This work has been supported in part by the Mitsubishi
 Foundation, the Grants-in-Aid for Scientific Research (14540271,
 17540275) and for Specially Promoted Research (13002001) of the
 Ministry of Education, Science, Sports and Culture of Japan.  Work at the University of Notre Dame was supported
by the U.S. Department of Energy under Nuclear Theory Grant DE-FG02-95-ER40934.
\end{acknowledgments}



\end{document}